# From Molecular to Metallic Gold Nanoparticles: The Role of Nanocrystal Symmetry in the Crossover Region

Sami Malola[1], Sami Kaappa[1] and Hannu Häkkinen[1,2,*]

Department of Physics, Nanoscience Center, University of Jyväskylä, FI-40014 Jyväskylä, Finland
Department of Chemistry, Nanoscience Center, University of Jyväskylä, FI-40014 Jyväskylä, Finland
* hannu.j.hakkinen@jyu.fi

**Popular summary:**

**Bulk gold is a good metal, *i.e.*, conductor of electricity and heat, due to its delocalized electron density that can respond to extremely small external perturbations such as electric field or temperature gradient. In energy space, the quantum states of the conduction electrons cross over the metal's Fermi energy continuously. But when gold is dispersed in finite nanometer-size particles or "clusters", the delocalized electronic states are re-grouped in energy space to "shells" according to allowed energy levels and the symmetry and shape of the atomic arrangement. This re-grouping generates also an energy gap in the vicinity of the Fermi energy akin to the energy gap between occupied and unoccupied electron orbitals in molecules. How does the formation of these shells and the energy gap affect the physico-chemical properties of the nanoparticle? How does one get from a "molecule-like" gold nanoparticle to "metal-like" gold nanoparticle?**

**Here we analyse the electronic structure and optical and chiroptical properties of recently reported gold nanoparticles of 144, 146, and 246 gold atoms, that are made by wet-chemistry methods and whose structures have been resolved to atomic precision. We demonstrate computationally how re-grouping of the quantum states of valence electrons can affect drastically the optical properties of nanoparticles in the crossover-size region, by either generating a multi-band "molecule-like" or a monotonous "plasmon-like" or "metallic" optical absorption. The lower the symmetry of the gold core, the more "metallic" is the nanoparticle. The underlying mechanism arises from symmetry-sensitive distribution of the electronic levels of the nanoparticle close to Fermi energy. Overall, this work sheds lights on fundamental mechanisms on how "molecular" nanoparticle properties can change "metallic" in the crossover-size region. Practical interest in these mechanisms derives from the need to design ambient-stable gold nanoparticles with tailored physico-chemical properties for applications in several areas such as plasmonics, nanomedicine, catalysis, biological imaging, sensing, and nanoelectronics**

**Abstract** Electronic structure and optical properties of the exceptionally stable, thiol-stabilized $Au_{144}(SR)_{60}$ nanoparticle (SR = SH, SPh, $SCH_2Ph$) are analyzed by using time-dependent density functional theory and density functional perturbation theory on atomistic models that are based both on the predicted chiral icosahedral (I) structure of the $Au_{144}S_{60}$ framework (Lopez-Acevedo *et al.*, J. Phys. Chem. 113, 5035 (2009)) and the recently (2018) confirmed crystal structure of icosahedral $Au_{144}(SCH_2Ph)_{60}$. Frontier orbitals of the nanoparticle, analyzed in the $I_h$ point group symmetry, show strong splitting by the order of 1 eV for the L = 4, L=5, and L = 6 spherical orbitals, manifesting a strong effect of the $Au_{144}$ nanocrystal on the electronic structure. Details of the splitting are almost unaffected by the choice of the ligand as long as the overall symmetry of the nanoparticle is kept, which confirms the dominant effect of the symmetry of the metal core on the electronic structure. Linear optical absorption exhibits non-plasmonic "molecular" behavior as previously extensively discussed in the literature, but now here analyzed in the context of the proper point group symmetry. We show that the choice of ligand only induces a slight red-shift of the absorption spectrum in the visible – near-IR region by going from SH to $SCH_2Ph$, which can be qualitatively understood by a spill-out of the metal electron density to the ligand layer by the larger ligand. The computed circular dichroism (CD) absorption spectrum in the region from 300 nm to 1100 nm shows very strong peaks that are up to 10 times stronger than those in the previously well-characterized chiral gold nanoparticle $Au_{38}(SR)_{24}$. This fact predicts promising chiroptical properties of this very stable nanoparticle in case it can be separated or made in enantiomer-pure form. Comparison of the electronic structure and optical properties of I – $Au_{144}(SR)_{60}$ to other well-known thiol-stabilized gold nanoparticles of similar size but of lower core symmetry, such as $C_{2v}$ – $Au_{146}(SR)_{57}$ and $C_1$ – $Au_{144}(SR)_{40}$, shows that the "molecular" features observed only in the linear absorption spectrum of I – $Au_{144}(SR)_{60}$ indeed originate from the high symmetry of this nanoparticle. Lower-symmetry particles have otherwise featureless linear absorption but show a weak plasmon-type peak signaling the approach of the metallic behavior. The effects of the discrete electronic shell structure on the non-plasmonic absorption of the well-known larger, $D_5$-symmetric $Au_{246}(SR)_{80}$ nanoparticle, are also analyzed here for the first time and reaffirm the notion about the drastic influence of point group symmetry of the metal core on the optical absorption of gold nanoparticles at the critical size range where transformation from "molecules" to "metals" takes place.

## I. INTRODUCTION

Gold nanoparticles are widely investigated for various applications in nanotechnology in areas such as plasmonics, nanomedicine, catalysis, biological imaging, sensing, and nanoelectronics [1-3]. Although it has been known since Faraday's times that gold can be dispersed in colloidal solutions as fine nanometer-size particles that display a fascinating spectrum of colors [4], it is only during the last decade that detailed investigations into structure-property relationship of gold nanoparticles have been made possible by great advances in synthesis, purification, and characterization of specific compounds with molecular formulas $Au_xL_y$ where L is the protecting ligand molecule, usually thiolate, phosphine, halide or alkynyl [5]. These nanoparticles are often called "monolayer protected clusters" since their total structure, including that of the ligand layer, can be solved to atomic precision by X-ray diffraction from samples where the clusters have formed high-quality single crystals. Largest currently known gold-based MPCs have close to 300 gold atoms and metal diameters of close to 3 nm [6-8].

Bulk gold is a simple monovalent s-metal and a good electric and thermal conductor. Confining the electrons in the delocalized Au(6s)-derived conduction band of bulk gold into a nanometer-size volume causes discretization of the electron states into a shell structure whose details depend on the size, shape and point group symmetry of the confining gold nanocrystal. These strong finite-size effects have for long been recognized to be an important factor in defining many physico-chemical properties of both bare and ligand-protected gold nanoclusters [9-13]. Of special interest is the energy gap that can form at the Fermi level. Simple considerations following ideas of Fröhlich [14] and Kubo [15] show that a three-dimensional electron gas confined in a volume V with N electrons has a density of states D at Fermi level $E_F$ such that $D(E_F) = 3N / 2E_F$ [12,16]. Thus, an average value of the energy gap $\Delta$ at the Fermi level is $\Delta(E_F) = 1/D(E_F) = 2E_F / 3N$. The existence and value of $\Delta(E_F)$ is relevant for considerations of the onset size for the "metallic" behavior of gold nanoparticles, manifested, e.g., by appearance of the surface plasmon resonance (SPR) or metallic electrochemical charging and redox properties. At room temperature, thermal excitations of electrons over the $\Delta(E_F)$ become relevant when $\Delta(E_F) \approx 25$ meV. Taking into account the Fermi energy of gold (5.5 eV) this leads to the onset size of 147 gold atoms and a spherical diameter of 1.7 nm [12]. One has to note that this average result is however strongly modified by the details of the electron shell structure which makes the level spacing highly non-uniform and can change the magnitude and location of $\Delta$ close to $E_F$.

Interestingly, it has been known since late 1990's that ambient-stable thiol-protected gold nanoparticles of size very close to the above-mentioned onset can indeed be abundantly synthesized both in organic and aqueous solvents [17]. First, they were referred to as "29 kDa" particles based on the approximate molecular mass. Refined mass measurements yielded first assignments for molecular formula as $Au_{144}(SR)_{59}$ [18], $Au_{144}(SR)_{60}$ [19], and $Au_{144\text{-}146}(SR)_{59\text{-}60}$ [20]. Their optical and electrochemical properties have since been extensively experimentally studied and computational models have been built for calculations of their electronic structure and optical properties [21-39]. The optical properties of these nanoparticles are known to indicate "non-metallic" behavior (i.e., absence of the SPR peak but exhibiting instead several absorption bands) [18-20, 22, 25, 26-29],
but on the other hand, electrochemical charging properties at room temperature indicate absence of $\Delta(E_F)$ and show an evenly spaced, multiple series of reduction-oxidation states [22-24] indicating a "metallic" behavior. Furthermore, the optical absorption spectra are known to be sensitive to temperature [28,29], to some extent to the nature of the ligand layer [36-38], and to elemental composition of the metal core, i.e., inter-metallic $Au_{144-x}Ag_x(SR)_{60}$ [31-33] and $Au_{144-x}Cu_x(SR)_{60}$ [34,35] particles show UV-visible absorption markedly different from the pure-gold particle.

A theoretical model for this nanoparticle was presented concurrently with the first accurate mass spectrometry data in 2009 as the first prediction of the atomic structure of $Au_{144}(SR)_{60}$ by Lopez-Acevedo *et al.* [21]. In that model, the gold core was built from concentric Mackay icosahedral 12- and 42-atom shells completed by an anti-Mackay shell of 60 gold atoms. This 114-atom structure was covered by 30 RS-Au-SR units (Supplemental Material, Fig. S1). The arrangement of these units makes the overall $Au_{144}S_{60}$ framework of chiral icosahedral (I) symmetry, that is also reflected in the slightly chiral arrangements of the 60-atom anti-Mackay shell at the metal-ligand interface (Fig. S1; the closest non-chiral structure of the 60-atom layer would be $I_h$ rhombicosi-dodecadedron). Calculations based on the density functional theory (DFT) showed that this model gave an excellent agreement in comparison of the computed powder X-ray diffraction function to the experimentally measured one [17], predicted a distinct electronic shell structure around $E_F$ giving the lowest allowed optical transitions in the mid-infrared, and indicated a metallic charging in electrochemistry as known at the time [22-24]. Absence of the central atom was argued by considerations to relieve built-up strain in the icosahedral metal structure. All of the later computational work by others has been based on this model with refinements regarding the symmetry of the organic part of the ligand layer [29,30] and occupation of the metal sites for modelling Au-Ag or Au-Cu 144-atom particles [32, 33, 35].

The emergence of the definite experimental structure of $Au_{144}(SR)_{60}$ (with SR = phenyl methyl thiolate, $SCH_2Ph$) from the single-crystal X-ray diffraction in 2018 [40] gives now firm grounds to analyze the structure-property relations of this nanoparticle (Fig. 1) The experimental structure directly confirms the predicted chiral icosahedral I-$Au_{144}S_{60}$ framework structure by Lopez-Acevedo *et al.* [21] and allows now detailed DFT-based analyses of the interplay between symmetry, electronic structure, optical absorption and circular dichroism (CD). We show here that the choice of the organic ligand has a minor effect on the electronic shell structure of the metal core and on the UV-visible absorption as long as the overall point group symmetry of the nanoparticle remains the same. By comparing to two other recently atomically defined nanoparticles of similar size, namely $C_{2v}$ – $Au_{146}(SR)_{57}$ [41] and $C_1$ – $Au_{144}(SR)_{40}$ [42,43], we show that lowering of the symmetry changes drastically the electronic shell structure around $E_F$ leading to a more uniform density of electron states, more featureless UV—visible absorption and emergence of a weak plasmon-like absorption band. Further comparisons to the atom-precise $D_5$-symmetric nanoparticle $Au_{246}(SR)_{80}$ [6] shows the drastic role of the high point group symmetry leading to a distinct shell structure and non-plasmonic absorption of this larger particle, not examined before by DFT calculations. The decisive role of the point group symmetry of the metal core and its effects on the electronic shell structure of gold nanoparticles close to the onset of metallic behavior have not been clearly demonstrated before. This discussion is of fundamental interest since it sheds lights on mechanisms how “molecular” nanoparticle properties can change “metallic” and of practical interest since understanding these mechanism helps design of ambient-stable nanoparticles for applications.

## II. COMPUTATIONAL METHODS AND SYSTEMS

Our comparative analysis is based on the experimental X-ray structure [40] of $Au_{144}(SCH_2Ph)_{60}$ and two structures derived from it (Fig. 1). By keeping the heavy atoms Au and S in places determined by the experimental structure, we made two additional structures where the sulfurs are passivated either by a phenyl group or hydrogen. The S-C bonds and the phenyl structure in the former case and the S-H bonds in the latter case were optimized using the real-space grid-based DFT package GPAW [44]. We used the Perdew-Burke-Ernzerhof (PBE) exchange-correlation functional [45] and 0.2 Å grid spacing for wave functions and densities. The optimization was stopped when the residual forces were below 0.05 eV/Å. The GPAW projector setups include scalar-relativistic effects for gold.

We analyzed the electronic shell structure of the three nanoparticles in two ways, by projecting the electron density in the Kohn-Sham orbitals either to spherical harmonics [11] or point-group ($I_h$) symmetry representations [46] as described elsewhere. Linear optical absorption and circular dichroism (CD) spectra were calculated by using the linear-response time dependent DFT (LR-TDDFT) module in GPAW [47]. PBE functional was used for electron-electron interactions and the electron wave functions and densities were treated in a real-space grid with spacing of 0.25 Å. Decomposition of a selected number of optical transitions into single electron-hole (e-h) excitations in Kohn-Sham basis was done in the framework of the density functional perturbation theory by using the so-called transition contribution map (TCM) method [48] with transition dipole contributions shown (DTCM). CD spectra were analyzed by decomposing the e-h contributions to the rotatory strengths (RTCM) including both the transition dipole moment and magnetic moment [49]. Both DTCM and RTCM analyses are convenient and flexible tools to decompose contributions to linear and CD absorption by multiple ways, e.g., by classifying the e-h pairs according to global symmetries (approximate angular momenta or point group representations), local symmetries near the atoms (corresponding to analysis by "atomic orbitals"), atom types, or spatial regions of the nanoparticle (metal core, atom shell, ligand layer).

To compare our results between the highly symmetric $Au_{144}(SR)_{60}$ and recently reported lower-symmetry particles of similar size, namely $C_{2v}$ – $Au_{146}(SR)_{57}$ [41] and $C_1$ – $Au_{144}(SR)_{40}$ [42,43], we analysed the electronic structure and linear absorption spectra of these compounds using the published structural information. The $C_{2v}$ – $Au_{146}(SR)_{57}$ particle was analyzed in a tri-anionic form using the experimental atomic coordinates from the work of Vergara *et al.* [41]. The atomic structure of $C_1$ – $Au_{144}(SR)_{40}$ is based on our earlier work [43] where the structure of the $Au_{144}$ core is taken from the electron microscopy data of Azubel *et al.* [42] and the ligand layer has been built around the core as explained in ref. [43]. The electronic structure and optical absorption of a larger, but still rather symmetric ($D_5$) $Au_{246}(SR)_{80}$ particle was analyzed as well, based on the published X-ray structure [6] of $Au_{246}(SPhCH_3)_{80}$ from which a simplified $Au_{246}(SH)_{80}$ model was built by fixing the Au-S framework to the experimental structure and optimizing all 80 S-H bonds.

## III. $I_h$ SYMMETRY AND THE ELECTRONIC STRUCTURE OF $Au_{144}(SR)_{60}$

Although the proper symmetry group for the thiol-protected $Au_{144}(SR)_{60}$ nanoparticle (Fig. 1) is the chiral icosahedral (I) group, it is instructive to analyze the electronic structure in the context of the $I_h$ group. Selecting this symmetry group is also justified by the fact that we found all frontier orbitals (orbitals that lie in the range from 1 eV below to 2 eV above of the Fermi energy) to be described by very high weights (typically 80-90 %) by representations belonging to $I_h$ symmetry class. The number of electrons in delocalized orbitals of this nanoparticle, originating mainly from Au(6s) electrons, is 84 by taking into account the electron-withdrawing nature of the 60 thiolates (144-60=84) [11]. The spherical harmonics analysis shows the presence of 1G, 1H, 2D, 3S, 1I, 2F, 1J, and 3P symmetries, corresponding qualitatively to the spherical electron gas model [50] (Fig. 2, cf. a similar analysis was first done by Lopez-Acevedo *et al.* [21]). One may note that although the above-mentioned shells span electron numbers up to a high count of 168, their energy ordering deviates from the exactly spherical 3D electron gas model as expected [50]. Fig. 3 gives the $I_h$-projected symmetries of the $Au_{144}(SR)_{60}$ nanoparticles protected by SH and $SCH_2Ph$ ligands. The following observations can be made by comparing Figs. 2 and 3: (i) the sequence of the sub-shells with $I_h$ symmetries is exactly the same in both systems for orbitals in the energy region of $-1 < E_f < 2$ eV, (ii) particularly the 1H(22) symmetric shell (Fig. 2) is strongly split below and above $E_f$ into $T_{2u}(6)$, $T_{1u}(6)$, and $H_u(10)$ sub-shells, where the electron occupations are shown in parentheses, and (iii) in most cases, the inter-shell energy gaps are very much independent of the system. Significant deviations in the electronic structure between the SH- and $SCH_2Ph$-protected particles are seen in the energy regions of -3 eV $< \varepsilon <$ -2 eV and $\varepsilon >$ 2 eV states due to the existence of the occupied and unoccupied $\pi$-electron states of the phenyl ring in the $SCH_2Ph$-protected particle. Due to the similarity of the Kohn-Sham electronic structure between these nanoparticles, optical absorption in the UV-visible region can also be expected to be rather similar, as we will next discuss in detail.

## IV. LINEAR OPTICAL ABSORPTION

Fig. 4 compares the linear absorption spectra of the SH-, SPh-, and $SCH_2Ph$-protected $Au_{144}$ nanoparticles as computed from LR-TDDFT. Interestingly, all three spectra show very similar features in the visible region, only a systematic red-shift is observed when going from $Au_{144}(SH)_{60}$ to $Au_{144}(SCH_2Ph)_{60}$ (674 to 738 nm, 530 to 566 nm, and 434 to 461 nm for the first three clear low-energy peaks as marked by arrows in Fig. 4). This almost constant red-shift in the peak energies (0.14 to 0.16 eV) can be qualitatively understood while examining the radially averaged Kohn-Sham all-electron potential (Fig. 5). One can see that the potential in the gold core including the Au-S layer, that is, in the $Au_{144}S_{60}$ moiety (first four minima), is very much unaffected by the nature of the ligand layer, but the extent of SPh and $SCH_2Ph$ ligands smoothens the potential for radial distances > 9Å as compared to the SH-protected particle, thus allowing a spill-out of the electrons in the metal core to the ligand layer. This decreases the average electron density in the gold core inducing the observed red-shift of the absorption peaks by arguments from the simple Drude model of plasmons in electron gas.

A more detailed analysis of three marked peaks (Fig. 4a) in the absorption spectrum of $Au_{144}(SCH_2Ph)_{60}$ is shown in Fig. 6 in terms of the dipole transition contribution maps (DTCMS). The electronic density of states is expressed in two ways, either projecting to the $I_h$ symmetry representations (SPDOS) or to local atomic orbitals (APDOS). The diagonal dashed line shows the condition where electron-hole energy difference equals the absorption energy, *i.e.*, $\varepsilon_e - \varepsilon_h = hc / \lambda$. It can be clearly seen that all e-h contributions with $\varepsilon_e - \varepsilon_h < hc / \lambda$ contribute constructively to the transition dipole moment while all e-h contributions with $\varepsilon_e - \varepsilon_h > hc / \lambda$ contribute de-constructively screening the dipole moment. Up to the 461 nm (2.7 eV) peak, the absorption is dominantly determined by the atomic Au(sp), Au(d), and S(p) states, while the pi-electrons in the phenyl ring of the ligand do not contribute. This also explains why the absorption spectra in Fig. 4 are so similar for all the three considered systems in the area where $\lambda >$ 400 nm, confirmed also by the DTCM analysis of the $Au_{144}(SH)_{60}$ particle shown in Fig. S2. The differences in the ligand layer become dominating for shorter wavelengths where the pi-electrons contribute.

## V. CIRCULAR DICHROISM

As discussed in the Introduction, the structural prediction for $Au_{144}(SR)_{60}$ by Lopez-Acevedo *et al.* [21] implies chirality in the system. As Fig. S1 shows in detail, the underlying non-chiral $I_h$ rhombicosi-dodecahedron (an Archimedean solid) has 30 square facets that are then distorted by bonding of the 30 RS-Au-SR diagonally on top of each facet. The bonding direction determines the handedness of the chiral structure when examined locally around each of the 12 $C_5$ axes. In addition, all sulfurs are intrinsically chiral centers, since the electronic bonding environment at each sulfur has four non-equivalent directions [52]. We have previously predicted from computations that all particles based on the predicted $Au_{144}(SR)_{60}$ structure [21] should have detectable CD signal in the UV-visible region [33]. Comparison of the computed CD absorption spectrum between the experimental $Au_{144}(SCH_2Ph)_{60}$ particle, its SH-derived model, and the previously well-characterized chiral $Au_{38}(SCH_2CH_2Ph)_{24}$ nanoparticle is shown in Fig. 7. In line with the comparison for linear absorption, the CD signals for $Au_{144}(SCH_2Ph)_{60}$ and $Au_{144}(SH)_{60}$ are very similar in the region $\lambda > 400$ nm but have distinct differences in the range of 300 – 400 nm due to the different ligand layer. Altogether, 14 maxima (+) or minima (-) can be seen in the CD spectrum of $Au_{144}(SCH_2Ph)_{60}$. The strongest peaks are about 10 times stronger than peaks in the CD spectrum computed for the $Au_{38}(SCH_2CH_2Ph)_{24}$ particle [51] using its experimental structure [54,55].

Analysis of the rotational transition contribution maps (RTCM) for three selected CD signals of $Au_{144}(SCH_2Ph)_{60}$ at 464 nm, 689 nm, and 947 nm shows that the strongest contributions (negative in the case of the negative signal at 464 nm and positive for the positive signals at 689 and 947 nm) come from hole states created in the occupied DOS in a narrow energy range of $-1.5$ eV $< \varepsilon <$ $-1$ eV (Fig. 8). These states are primarily located at the 60-atom Au layer and at the 30-unit SR-Au-SR layer at the interface between the gold nanocrystal and the ligand layer. Thus, the effect of the chiral metal-ligand interface is directly manifested in the chiroptical behavior of $Au_{144}(SCH_2Ph)_{60}$.

## VI. COMPARISON OF I – $Au_{144}(SR)_{60}$ TO $C_1$ – $Au_{144}(SR)_{40}$, $C_{2v}$ – $Au_{146}(SR)_{57}$, AND $D_5$ – $Au_{246}(SR)_{80}$

Two different gold nanoparticles, closely related in composition and size to $Au_{144}(SR)_{60}$, have been recently reported. Vergara *et al.* published [41] the X-ray crystal structure of $Au_{146}(4MBA)_{57}$ particle and Azubel *et al.* reported [42] the atomic structure of the 144-atom gold core of the 3MBA protected particle where the mass spectrometry yielded a low number (about 40) of ligands, hence the formulation of $Au_{144}(3MBA)_{40}$ (4MBA/3MBA is para/meta mercaptobenzoic acid, respectively). The $Au_{146}(4MBA)_{57}$ has a $C_{2v}$-symmetric, fcc-based metal core with one stacking fault while the core of $Au_{144}(3MBA)_{40}$ has only $C_1$ symmetry (see insets to Fig. 9a,b). Since the count of metal free electrons is similar to $Au_{144}(SR)_{60}$ particle, it is interesting to compare the details of the electronic structure and its implication to optical absorption between these systems. As Figs. 9a,b show, the computed optical absorption spectra of both of these lower-symmetry nanoparticles is qualitatively different from $Au_{144}(SR)_{60}$ (Fig. 4), showing otherwise monotonous increase in absorption with decreasing wavelength but a single weak absorption band at 520 nm for $Au_{144}(3MBA)_{40}$ and 443 nm for $Au_{146}(4MBA)_{57}$. The DTCM analysis for those energies reveals that the absorption band is caused by massive contributions from single e-h transitions from occupied states to unoccupied states within $E_F \pm 2.5$ eV. In general, transitions where $\varepsilon_e - \varepsilon_h < hc / \lambda$ contribute constructively to the absorption peak while transitions where $\varepsilon_e - \varepsilon_h > hc / \lambda$ contribute via screening. In the case of $C_1$-symmetric $Au_{144}(3MBA)_{40}$, a separate constructive "diagonal band" by occupied and unoccupied states within $E_F \pm 1$eV is formed, caused by the high dispersion of states close to $E_F$, in other words, by disappearance of the clear electronic shell structure.

Jin and collaborators reported in 2016 [6] the crystal structure of a $D_5$-symmetric $Au_{246}(SPhCH_3)_{80}$ nanoparticle, which, interestingly, did not show a plasmonic optical spectrum, but instead showed several distinguishable absorption bands similar to the $Au_{144}(SR)_{60}$ case [56]. We analyzed here the electronic structure and optical absorption of the simplified model structure $Au_{246}(SH)_{80}$. As Fig. 10 shows, the linear optical absorption curve shows five distinguishable bands for $\lambda > 400$ nm. The electronic structure of this nanoparticle shows that the states close to $E_F$ are grouped to distinct shells according to $D_5$ symmetry representations [46] and the DTCM analysis of the 432 nm band in Fig. 9c shows that this shell structure leads to e-h contributions to this band that are qualitatively much different from the smaller, low symmetry particles in (a) and (b).

## VII. CONCLUSIONS

In this work we have systematically analyzed the electronic structure and optical properties of several recently reported gold nanoparticles between 144 and 246 gold atoms, that are made by wet chemistry and whose structures have been resolved to atomic precision. This size region is important for understanding of how the nanoparticles' properties change from "molecular" to "metallic". We demonstrate the dominant role of the point group symmetry of the gold nanocrystal in determining the electronic shell structure of the particle, which in turn determines the details in the linear optical absorption spectrum as well as the grouping of the states close to the Fermi energy, defining the location and magnitude of the fundamental energy gap distinct for "molecular" particles. For particles of 144 to 146 gold atoms, the lower point group symmetry of the gold nanocrystal smears out the electronic shells creating a more continuous density of states close to $E_F$, a more monotonous linear optical absorption, and development of a single plasmon-like absorption band. Identification of clear electronic shell structure as projected to the $D_5$ symmetry of the $Au_{246}(SR)_{80}$ explains the non-plasmonic absorption of this quite large nanoparticle that otherwise could be expected [7,28] to be already plasmonic. Interestingly, silver nanoparticles have been predicted [57,58] and observed [59,60] to have plasmonic absorption at smaller sizes (136 to 141 silver atoms) than gold particles. This qualitative difference is likely to arise from different screening properties of the metal d-electrons due to the clearly different locations of the Ag(4d) and Au(5d) derived bands with respect to the Fermi energy of the nanoparticle. Finally, our work predicts that in the case of a successful enantioseparation of $Au_{144}(SR)_{60}$, the handedness of the enantiopure particles can be undoubtedly recognized from the CD absorption spectrum facilitating their use in applications that require optical isomerism.

## ACKNOWLEDGEMENTS

This work was supported by the Academy of Finland (grant 294217 and H.H.'s Academy Professorship). S.K. thanks the Väisälä Foundation for a PhD study grant. Computations were made at the CSC supercomputer center in Espoo (Finland) and at the Barcelona supercomputing center (Spain) as part of the PRACE project NANOMETALS. We thank Zhikun Wu, Ignazio Garzon, Xochitl Lopez-Lozano, Hans-Christian Weissker, Alessandro Fortunelli, Sandra Vergara, Robert L. Whetten, Maia Azubel, and Roger D. Kornberg for numerous discussions on the structure and properties of the $Au_{144}(SR)_{60}$, $Au_{146}(SR)_{57}$ and $Au_{144}(SR)_{40}$ nanoparticles. We thank R.L. Whetten for providing a copy of manuscript [39] prior to publication.

## REFERENCES AND NOTES

* Corresponding author, email: hannu.j.hakkinen@jyu.fi

**FIGURES**

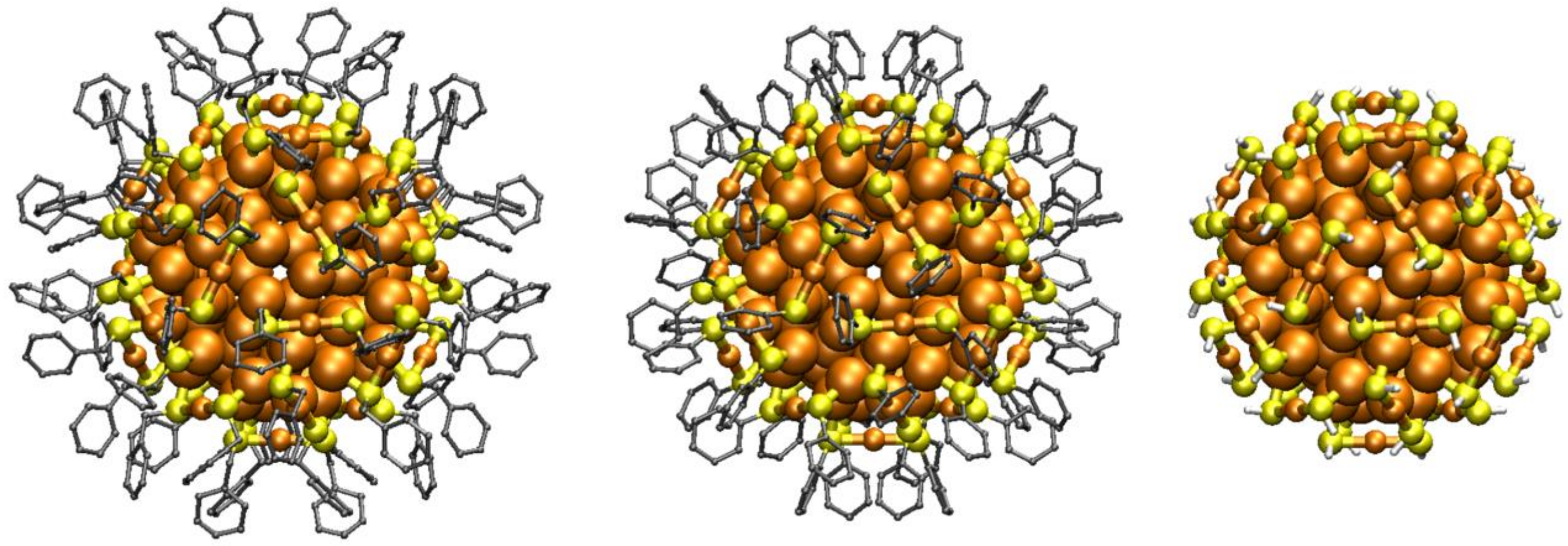

FIG. 1. Structures of the $Au_{144}(SR)_{60}$ nanoparticles considered in this work. Left: the experimental structure of $Au_{144}(SCH_2Ph)_{60}$ [40], center and right: simplified models based on the experimental structure but with ligands SPh (center) and SH (right). Au: orange, S: yellow, C: gray, H: white.

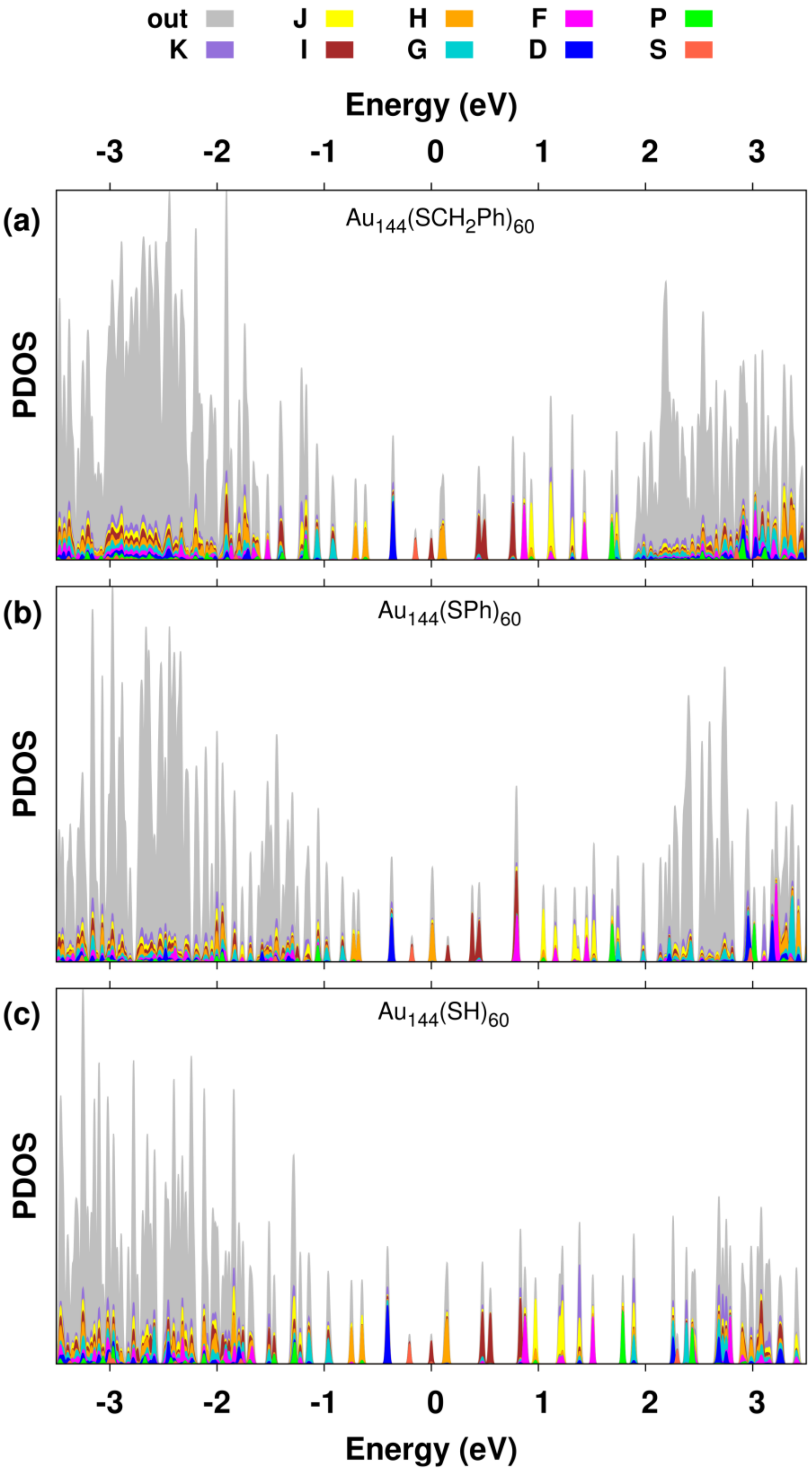

FIG. 2. Projection of the density of electron states (PDOS) to spherical harmonics for Kohn-Sham orbitals in the vicinity of the Fermi energy (E=0) for (a) $Au_{144}(SCH_2Ph)_{60}$ , (b) $Au_{144}(SPh)_{60}$, and (c) $Au_{144}(SH)_{60}$. Each of the states is broadened with a 0.01 eV gaussian.

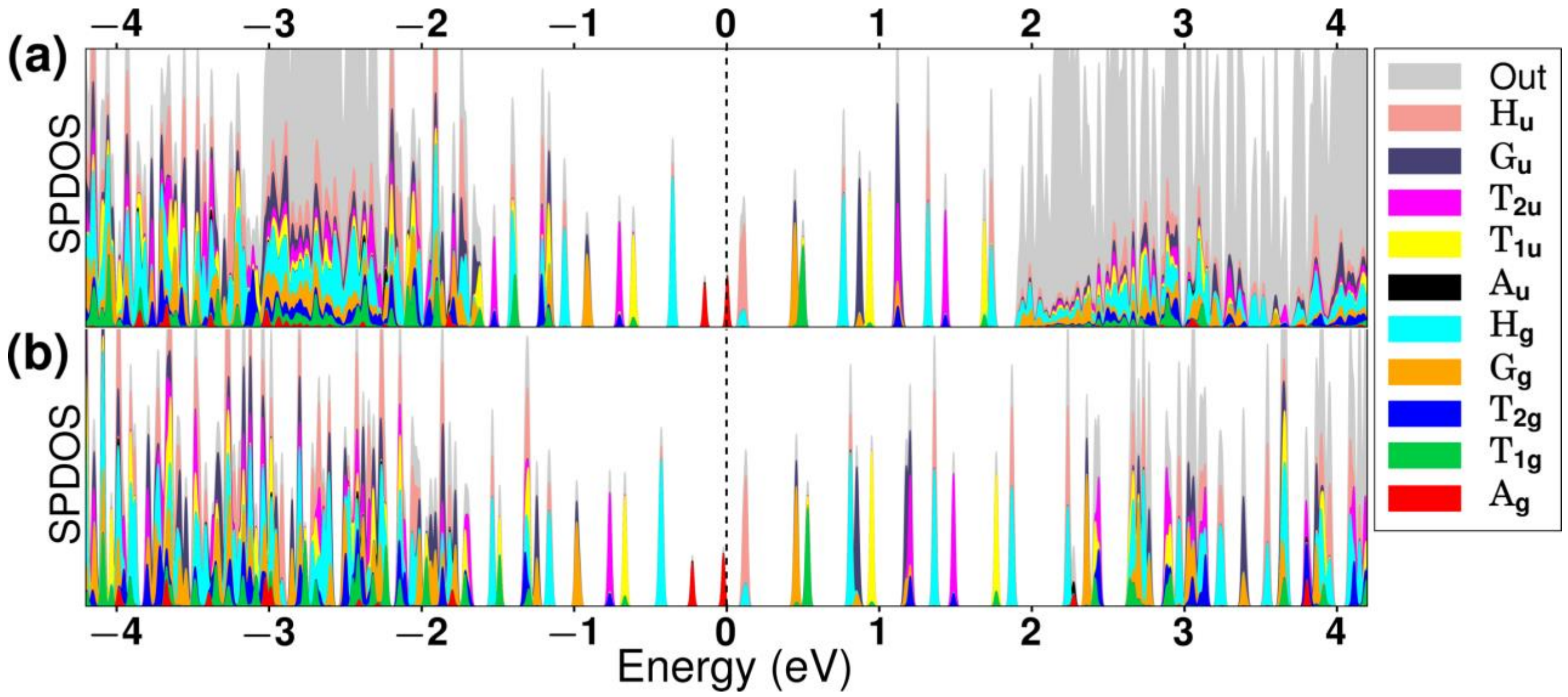

FIG. 3. $I_h$ symmetry-projected electron density of states (SPDOS) for (a) $Au_{144}(SCH_2Ph)_{60}$ and (b) $Au_{144}(SH)_{60}$ in the vicinity of the Fermi energy (E=0). Each of the states is broadened with a 0.01 eV gaussian.

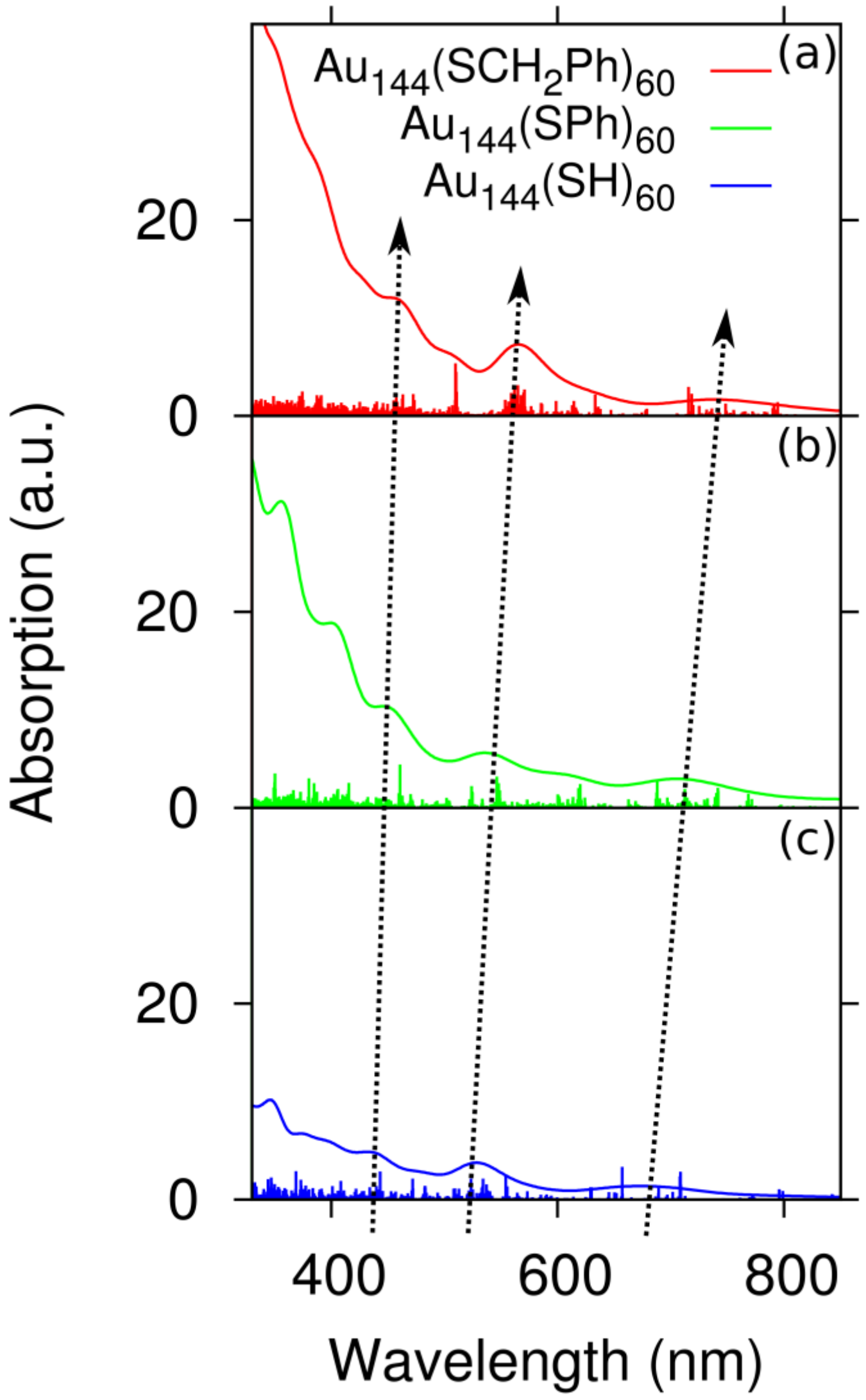

FIG. 4. Computed linear optical absorption spectra of (a) $Au_{144}(SCH_2CH_2Ph)_{60}$, (b) $Au_{144}(SPh)_{60}$ and (c) $Au_{144}(SH)_{60}$. The individual optical lines (marked by sticks) are broadened by 0.075 eV gaussians for the continuous curves. The red-shifts of the lowest three absorption peaks are shown by the dotted arrows.

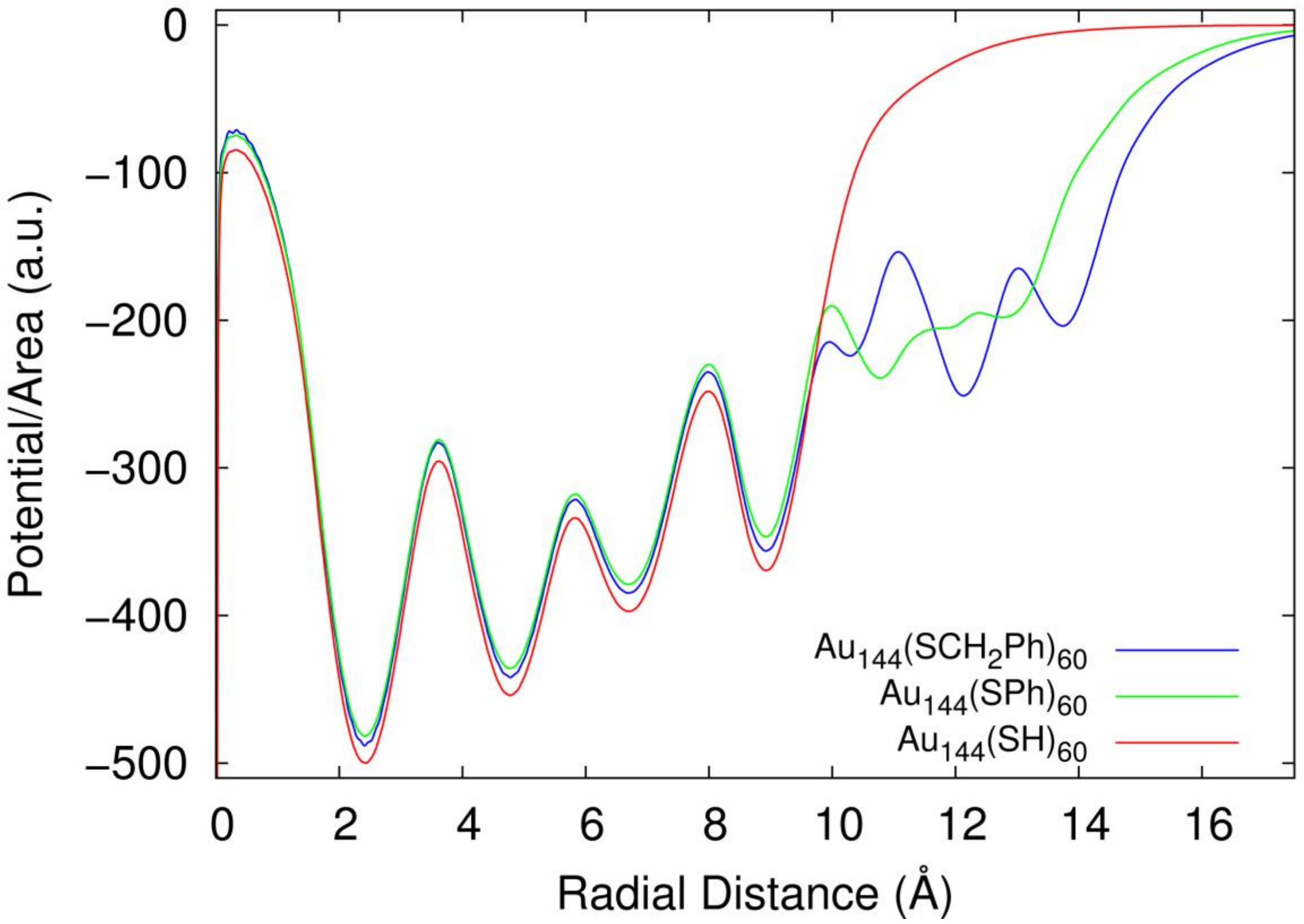

FIG. 5. Radially averaged total single-particle Kohn-Sham potential for $Au_{144}(SH)_{60}$, $Au_{144}(SPh)_{60}$ and $Au_{144}(SCH_2CH_2Ph)_{60}$

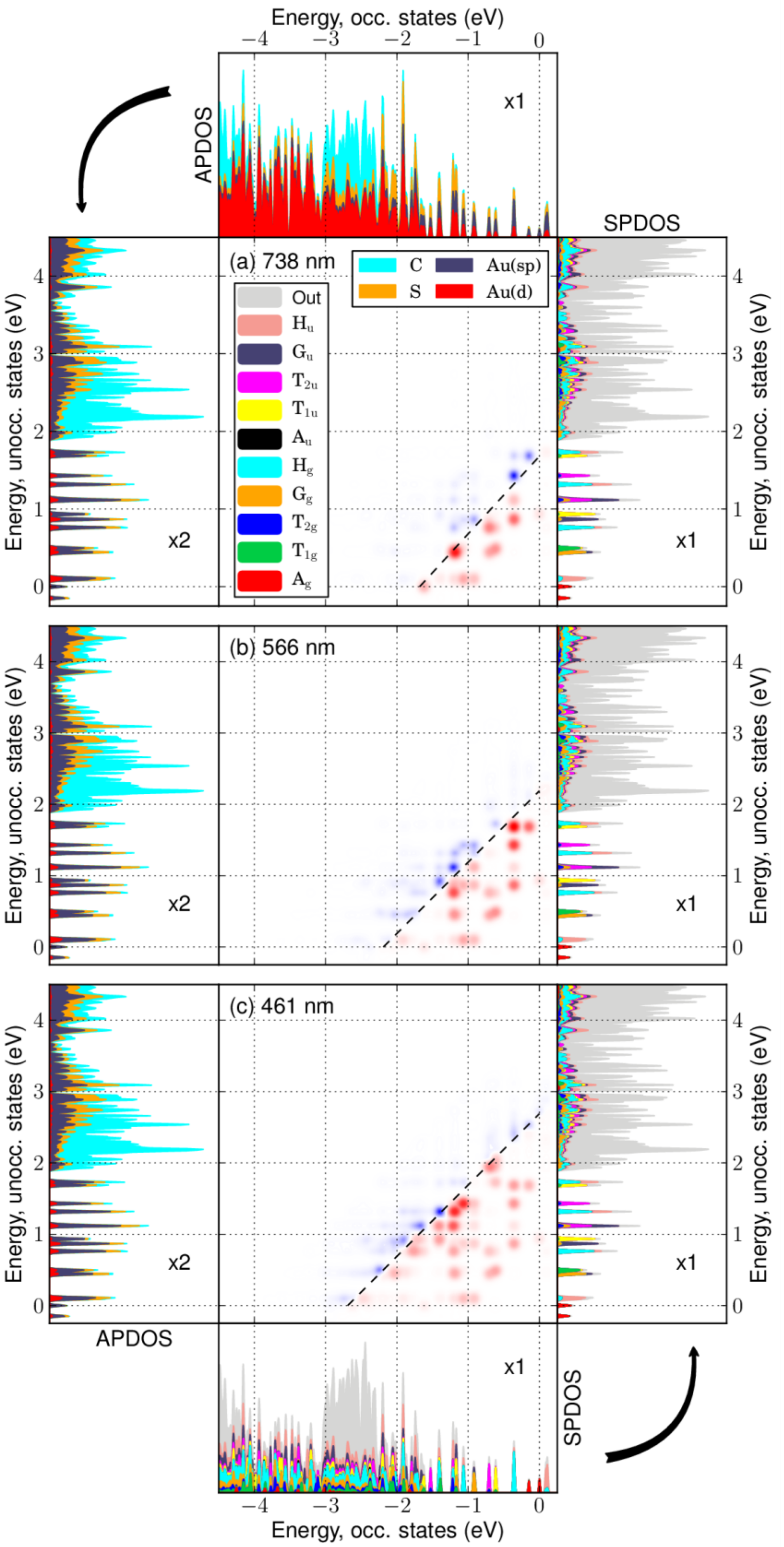

FIG. 6. DTCM analysis of the three lowest-energy linear absorption peaks of $Au_{144}(SCH_2Ph)_{60}$ as marked in Fig. 4(a). The panels (a)-(c) at the center show single-particle contributions to the absorption at the given wavelength (energy). Red/blue contributions denote constructive/destructive contribution to the transition dipole. The brightness of the red/blue spots scales with the magnitude of contribution. The dashed diagonal lines denote the electron-hole (e-h) energy equaling the peak position, i.e., $\varepsilon_e - \varepsilon_h = hc / \lambda$. The electron states are formed in the manifold of the initially unoccupied states (right and left panels in (a)-(c)) and the hole states are formed in the manifold of the initially occupied states (top and bottom panels). The occupied – unoccupied electron density of states (DOS) is presented in two alternative ways. Bottom-right panels show the decomposition of the DOS projected to $I_h$ symmetry representations (SPDOS) while the top-left panels show projection to spherical atomic orbitals (atom-projected APDOS). The Fermi energy is at zero.

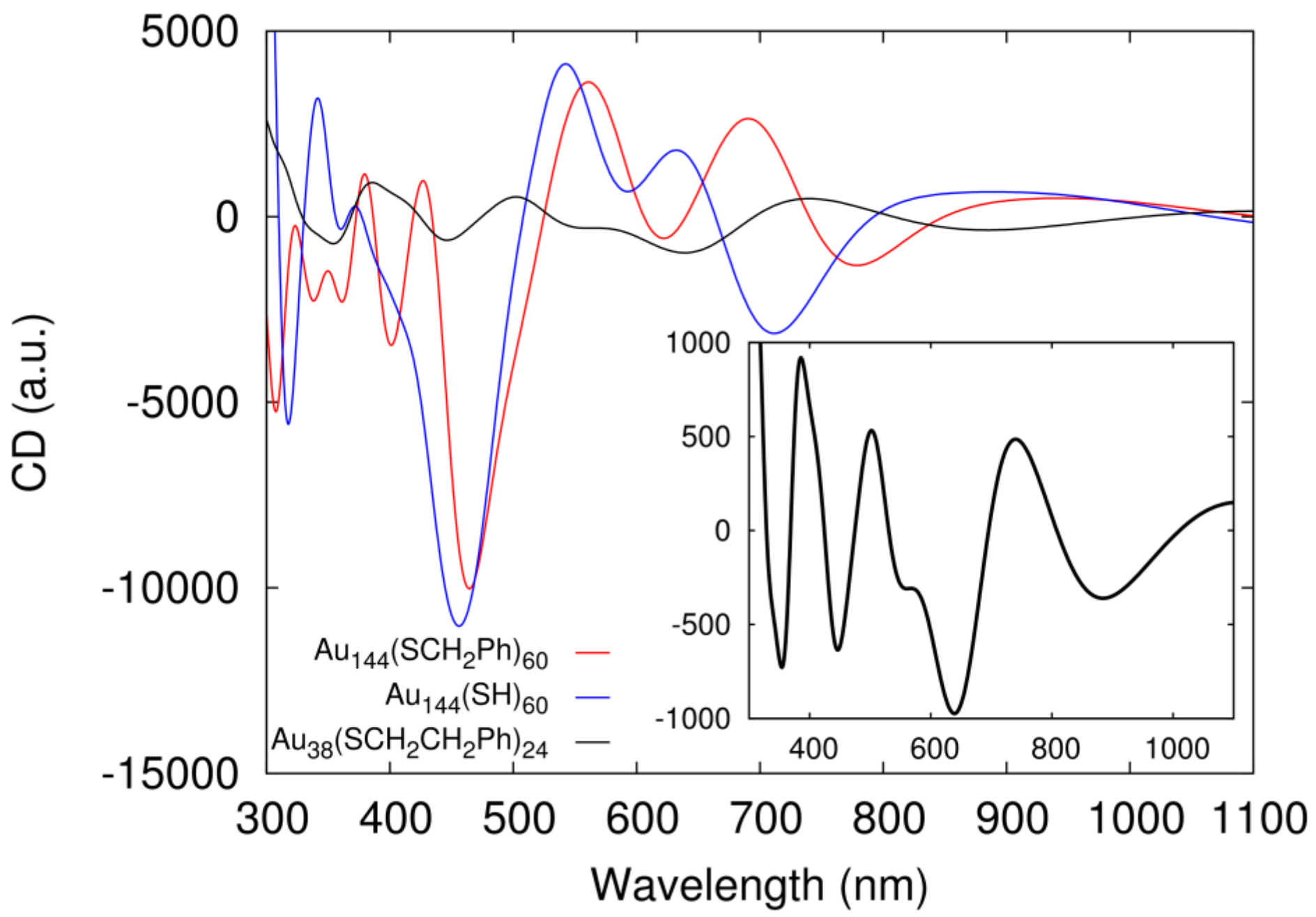

FIG. 7. Computed CD absorption spectra for $Au_{144}(SCH_2Ph)_{60}$ (red curve) and $Au_{144}(SH)_{60}$ (blue curve) as compared to the computed CD absorption spectrum of $Au_{38}(SCH_2CH_2Ph)_{24}$ (black curve) taken from ref. [51]. The inset shows the $Au_{38}(SCH_2CH_2Ph)_{24}$ data in a zoomed scale. Note the y-scales: the strongest peaks in the $Au_{144}(SR)_{60}$ spectra are almost 10 times higher than peaks at similar wavelengths in the $Au_{38}(SR)_{24}$ spectrum. The magnitudes of the computed CD signals for all particles are directly comparable since the absorption cross sections show absolute values per one particle. The CD spectrum for $Au_{144}(SCH_2Ph)_{60}$ and $Au_{144}(SH)_{60}$ was calculated for the “right-handed” enantiomers shown in Figs. 1a and 1c, where the arrangements of the 5 RS-Au-SR units around each $C_5$ axis forms a “right-down” blade structure.

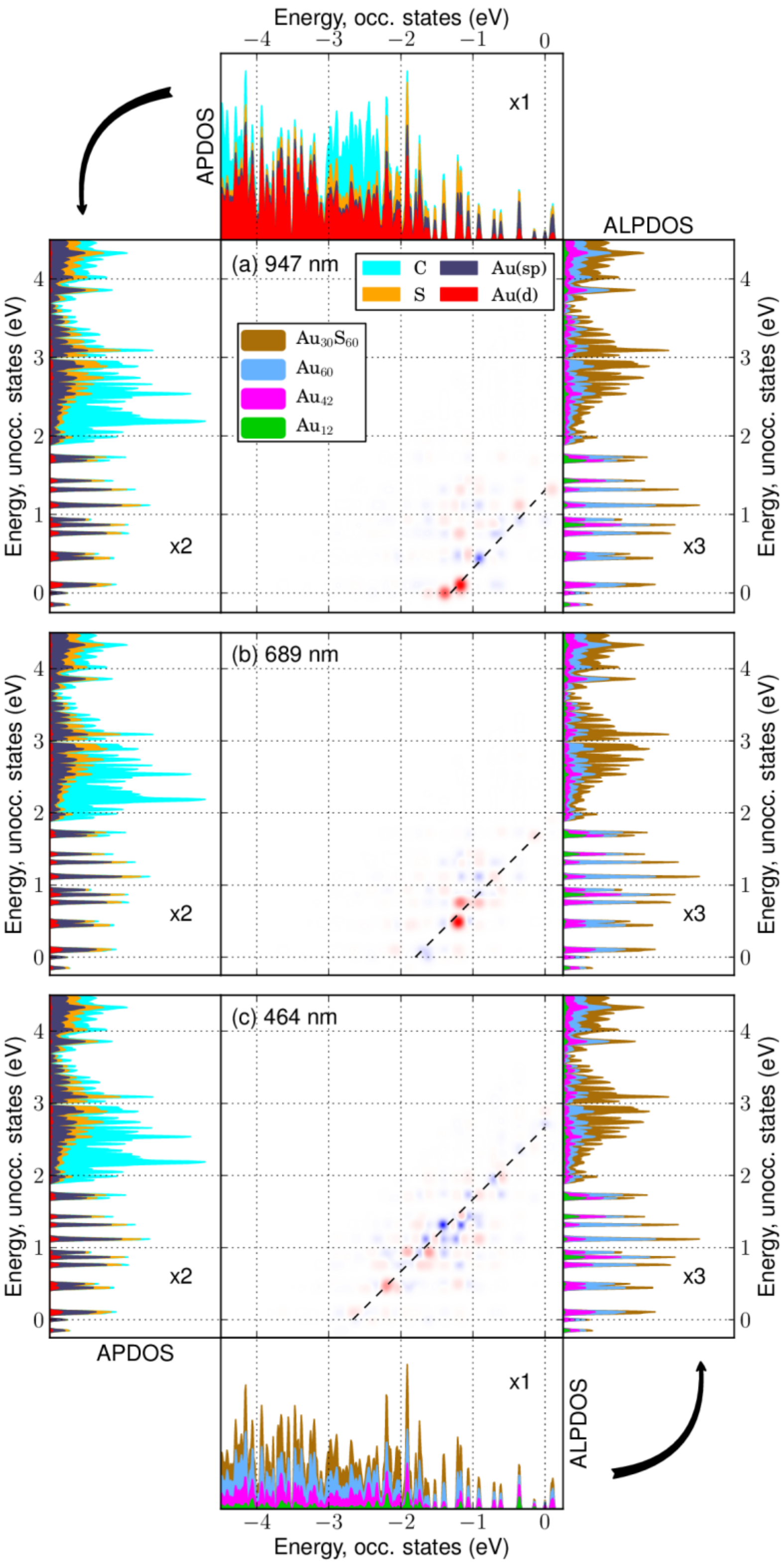

FIG. 8. RTCM analysis of three selected peaks in the CD absorption spectrum of $Au_{144}(SCH_2Ph)_{60}$, shown in Fig. 7. The layout is similar to Fig. 6, but here the red/blue spots denote constructive/destructive contributions to the rotational strengths. The single electron DOS is presented either as projected to local atomic orbitals (APDOS, bottom and right panels) or by layer-by-layer manner (ALPDOS, top and left panels). The dashed diagonal lines denote the electron-hole (e-h) energy equaling the peak position ($\varepsilon_e$ - $\varepsilon_h$ = hc / λ). The Fermi energy is at zero

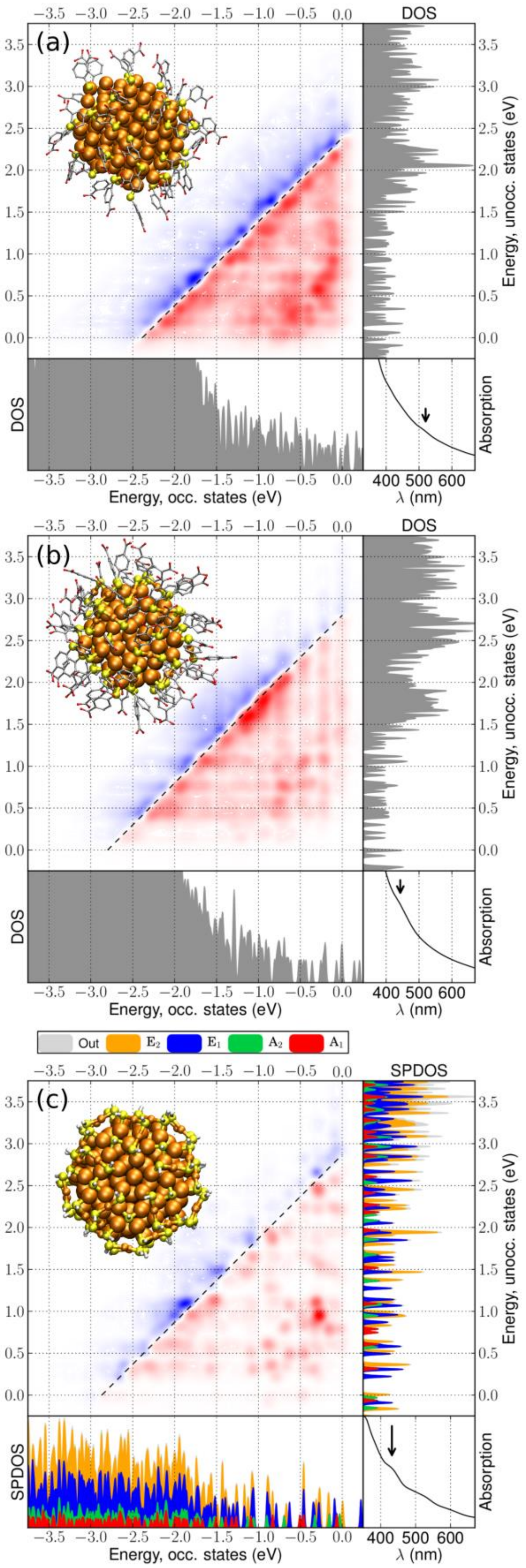

FIG. 9. DTCM analysis of the computed linear absorption spectrum of (a) $Au_{144}(3MBA)_{40}$ , (b) $[Au_{146}(4MBA)_{57}]^{3-}$ and (c) $Au_{246}(SH)_{80}$, with nanoparticle structures shown as insets. The panels in the lower right corner show the computed absorption spectra, and the arrows show the analyzed absorption energy (520 nm in (a), 443 nm in (b), and 432 nm in (c)). The layout is otherwise similar to Fig. 6, but the occupied and unoccupied electronic density of states (DOS) is shown in (a) and (b) without any symmetry assignments, and with $D_5$ representations in (c). The structures of the nanoparticles are shown as insets. The dashed diagonal lines denote the electron-hole (e-h) energy equaling the peak position ($\varepsilon_e - \varepsilon_h = hc / \lambda$). The Fermi energy is at zero. Note the qualitative changes in the density of electron states close to $E_F$ when going from (a) to (c), with the DOS being most continuous in (a) and most peaked (to shells with well-defined $D_5$ representations) in (c).

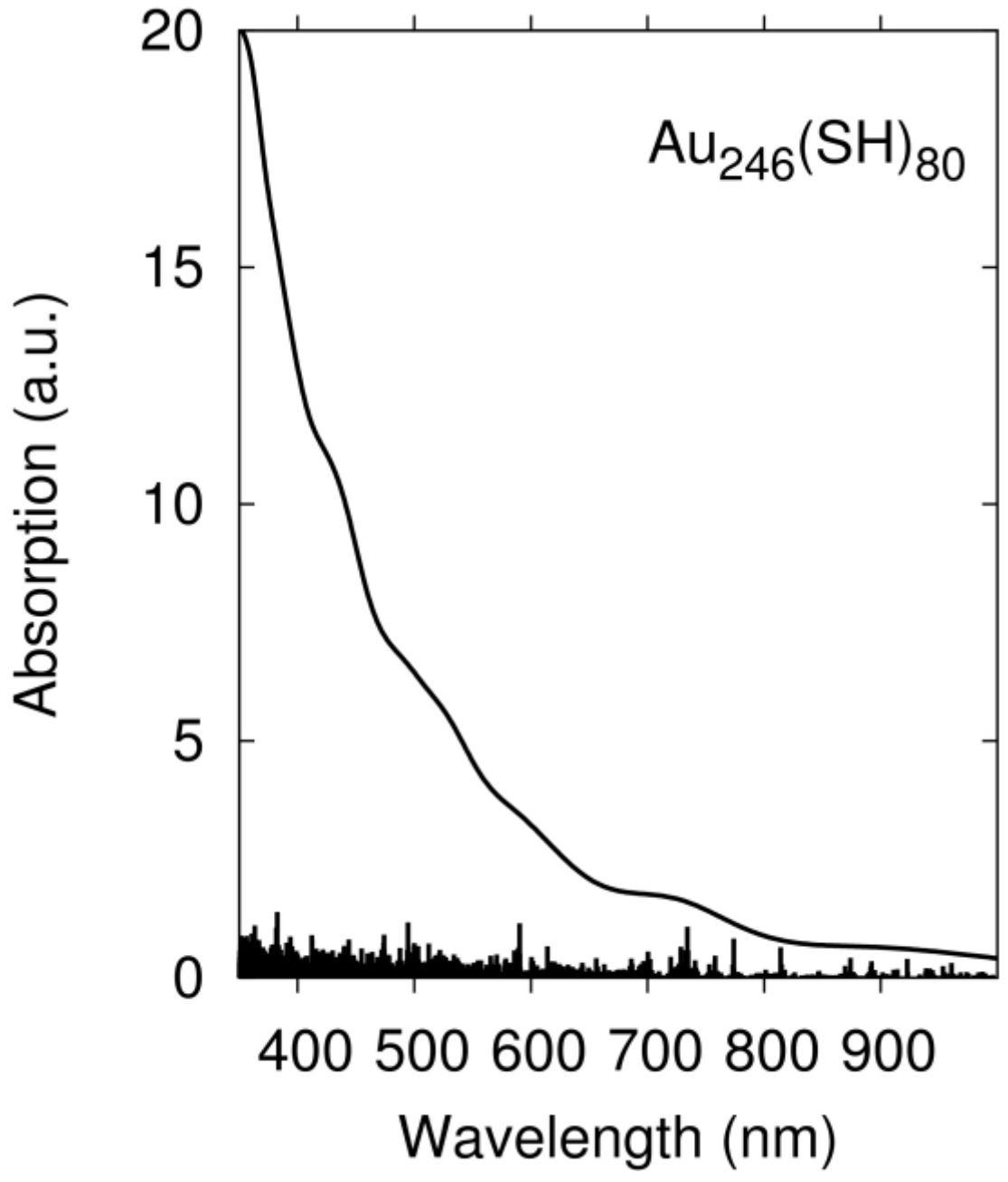

FIG. 10. Computed linear optical absorption spectrum of $Au_{246}(SH)_{80}$. The individual optical lines (marked by sticks) are broadened by 0.075 eV gaussians for the continuous curves. Clear absorption bands are observed at 432 nm, 516 nm, 590 nm, 713 nm, and 897 nm, agreeing qualitatively well with the experiment on $Au_{246}(SPhCH_3)_{80}$ [56].

**SUPPLEMENTAL MATERIAL**

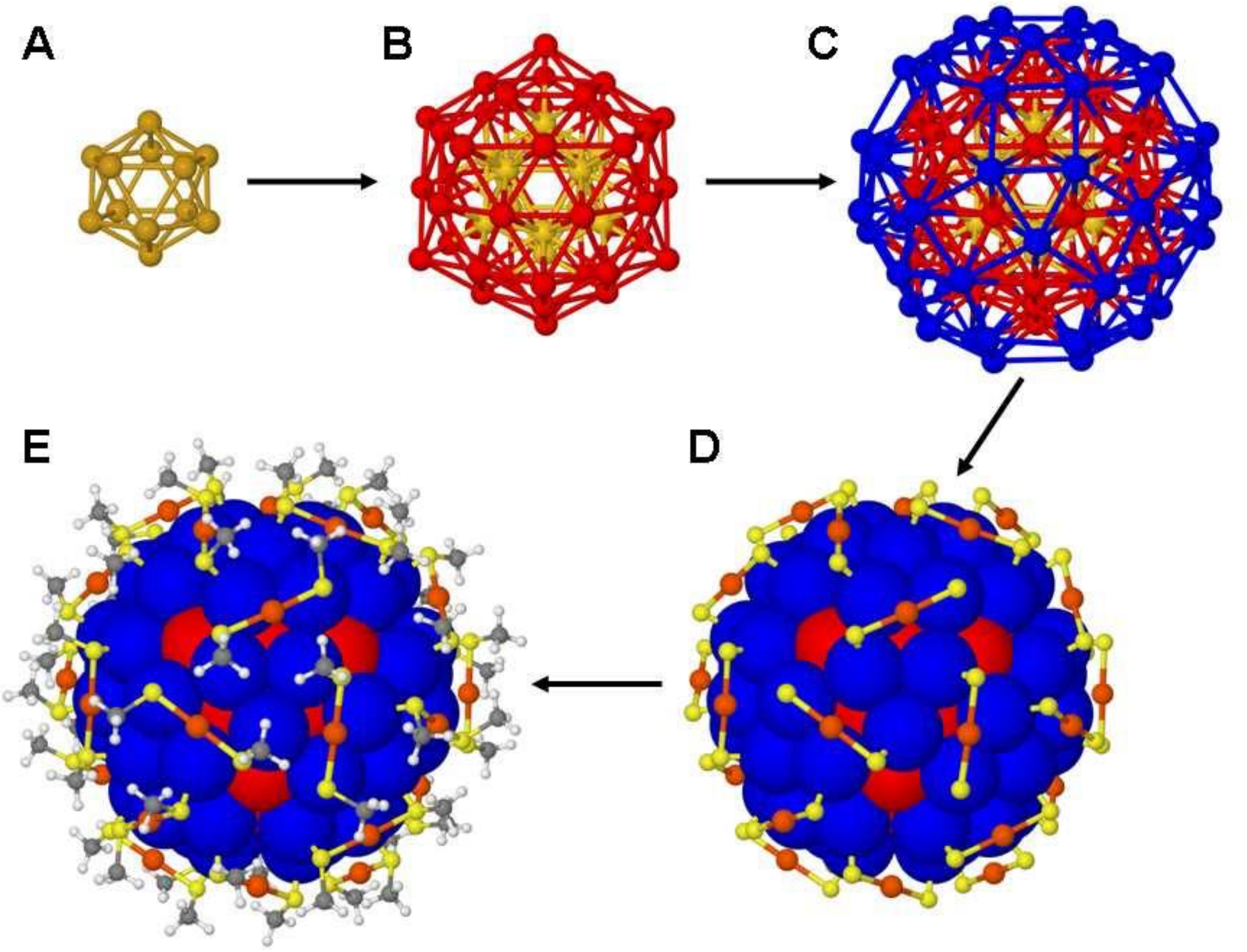

FIG. S1. Shell-by-shell visualization of the predicted structural model for $Au_{144}(SR)_{60}$ by Lopez-Acevedo *et al.* [21]. (A)-(C) show the first two Mackay (12 + 42 atoms) and the third anti-Mackay shell (60 atoms) of the gold core. The 60-atom shell forms a chiral structure where the initially 30 square-like four-atom facets of the corresponding non-chiral structure ($I_h$ rhombicosi-dodecahedron) are distorted to rhombus-like by bonding of the 30 RS-Au-SR units on top of the 60-atom layer (D). However, the chiral distortion of the 60 atom layer does not reach the limit of I-snub dodecahedron, where the rhombus-like units would compose of two equilateral triangle with a joint edge. (E) shows the ligand layer built from small model thiolates $SCH_3$. Reproduced from ref. [21] by permission. Copyright 2009 American Chemical Society.

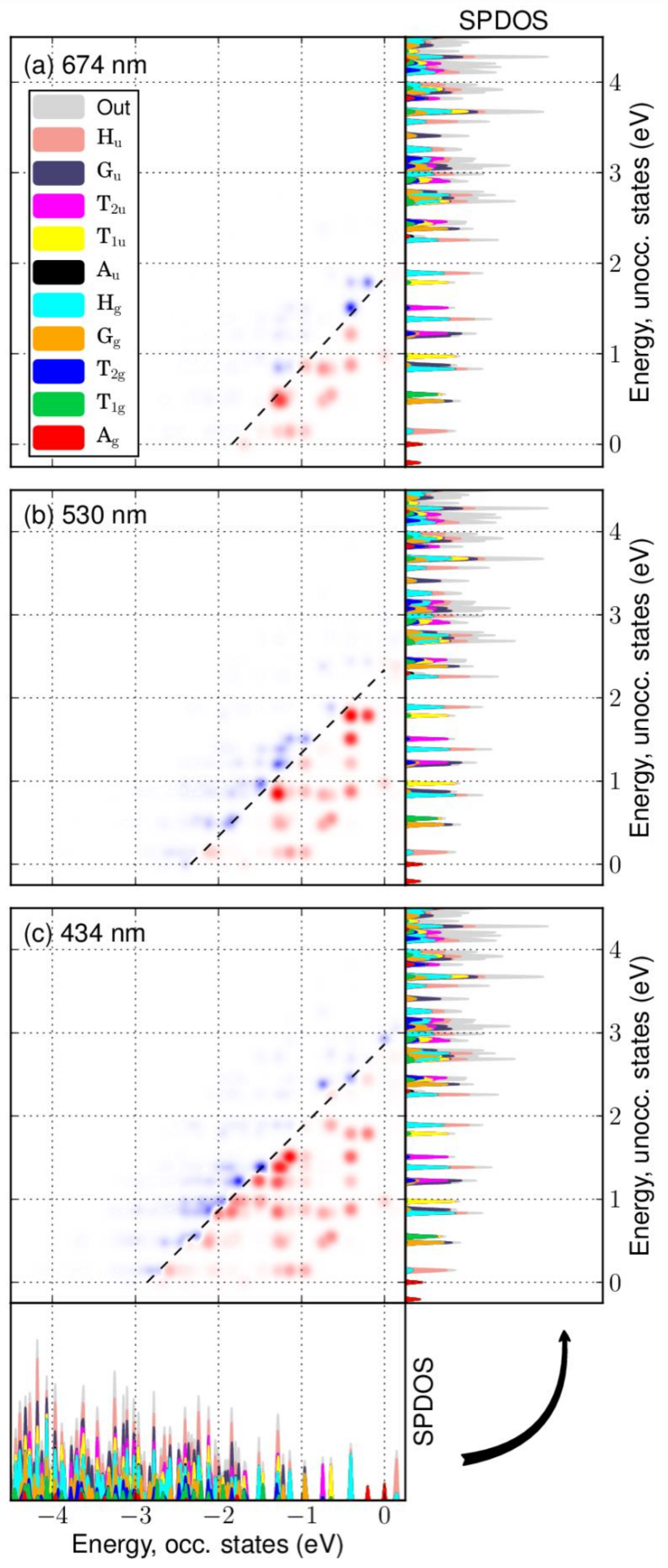

FIG. S2. DTCM analysis for the three lowest-energy absorption peaks (Fig. 4 in the main text) for the $Au_{144}(SH)_{60}$ nanocluster. Note that the e-h constructive/destructive contributions to the peaks are almost identical to the ones shown in Fig. 6 for $Au_{144}(SCH_2Ph)_{60}$. The dashed diagonal lines denote the electron-hole (e-h) energy equaling the peak position ($\varepsilon_e$ - $\varepsilon_h$ = hc / λ). The Fermi energy is at zero.